\documentclass[twocolumn,tighten]{aastex62}

\usepackage{xcolor}
\usepackage{hyperref}
\usepackage[caption=false]{subfig}
\usepackage{longtable}
\usepackage{afterpage}
\usepackage{xspace}
\usepackage{upgreek}
\usepackage{multirow}

\graphicspath{{./}{figures/}}

\shorttitle{Narrow-Band Signal Localization for SETI}
\shortauthors{Brzycki et. al.}

\begin{document}

\newcommand{\tseti}{\textsc{\emph{turbo}SETI} }
\newcommand{\tiddalik}{\textsc{Tiddalik} }
\newcommand{\blimpy}{\textsc{Blimpy} }
\newcommand{\Hzs}{Hz\,s$^{-1}$}
\newcommand{\EventBW}{$\Delta\nu_{\rm{event}}$}

\newcommand{\nstar}{{\color{black}1327 }}
\newcommand{\nstaradd}{{\color{black}641 }}  

\newcommand{\nobsgbtS}{{\color{black}6456 }}
\newcommand{\nobsgbtL}{{\color{black}6042 }}
\newcommand{\nobsgbtSL}{{\color{black}12504 }}

\newcommand{\nstargbt}{{\color{black}1138 }}
\newcommand{\nstargbtS}{{\color{black}1005 }}
\newcommand{\nstargbtL}{{\color{black}882 }}
\newcommand{\nstargbtSL}{{\color{black}749 }}

\newcommand{\ncadencegbt}{{\color{black}2089 }}
\newcommand{\ncadencegbtS}{{\color{black}1076 }}
\newcommand{\ncadencegbtL}{{\color{black}1013 }}

\newcommand{\nhrpks}{483.0}
\newcommand{\nhrgbtL}{506.5}
\newcommand{\nhrgbtS}{538.0}

\newcommand{\nobspks}{{\color{black}5796}\xspace}
\newcommand{\nstarpks}{{\color{black}189}\xspace}
\newcommand{\nstarpksIsaacson}{{\color{black}183}\xspace}
\newcommand{\nstarpksExtra}{{\color{black}6}\xspace}
\newcommand{\ncadencepksdb}{{\color{black}990}\xspace}
\newcommand{\ncadencepks}{{\color{black}966}\xspace}

\newcommand{\nevent}{{\color{black} zero candidates}\xspace}
\newcommand{\neventpks}{{\color{black}77\xspace}}
\newcommand{\neventgbtL}{{\color{black}15998}\xspace}
\newcommand{\neventgbtS}{{\color{black}5102}\xspace}

\newcommand{\neventpksnstar}{20\xspace}  
\newcommand{\neventgbtLnstar}{831\xspace}
\newcommand{\neventgbtSnstar}{511\xspace}

\newcommand{\ngroupspks}{{\color{black}60}\xspace}
\newcommand{\ngroupsgbtL}{{\color{black}4522}\xspace}
\newcommand{\ngroupsgbtS}{{\color{black}1572}\xspace}

\newcommand{\nfinalpks}{0\xspace}
\newcommand{\nfinalL}{0\xspace}
\newcommand{\nfinalS}{0\xspace}

\newcommand{\nhitspks}{4.45M\xspace}
\newcommand{\nhitsgbtL}{{\color{black}37.14M}\xspace}
\newcommand{\nhitsgbtS}{{\color{black}10.12M}\xspace}

\newcommand{\nhitsgbtLzerodrift}{21.90M\xspace}
\newcommand{\nhitsgbtLnegativedrift}{13.9M\xspace}
\newcommand{\nhitsgbtLpositivedrift}{1.37M\xspace}

\newcommand{\nhitsgbtSzerodrift}{7.36M\xspace}
\newcommand{\nhitsgbtSpositivedrift}{1.21M\xspace}
\newcommand{\nhitsgbtSnegativedrift}{1.55M\xspace}


\newcommand{\BI}{\textit{Breakthrough Initiatives} }
\newcommand{\BLI}{\textit{Breakthrough Listen Initiative} }
\newcommand{\BL}{BL\xspace}

\title{Narrow-Band Signal Localization for SETI on Noisy Synthetic Spectrogram Data}


\newcommand{\UCB}{Department of Astronomy,  University of California Berkeley, Berkeley CA 94720}
\newcommand{\SSL}{Space Sciences Laboratory, University of California, Berkeley, Berkeley CA 94720}
\newcommand{\SWIN}{Centre for Astrophysics \& Supercomputing, Swinburne University of Technology, Hawthorn, VIC 3122, Australia}
\newcommand{\GBT}{Green Bank Observatory,  West Virginia, 24944, USA}
\newcommand{\OXF}{Astronomy Department, University of Oxford, Keble Rd, Oxford, OX13RH, United Kingdom}
\newcommand{\NIJ}{Department of Astrophysics/IMAPP,Radboud University, Nijmegen, Netherlands}
\newcommand{\ATNF}{Australia Telescope National Facility, CSIRO, PO Box 76, Epping, NSW 1710, Australia}
\newcommand{\HOU}{Hellenic Open University, School of Science \& Technology, Parodos Aristotelous, Perivola Patron, Greece}

 \newcommand{\USQ}{University of Southern Queensland, Toowoomba, QLD 4350, Australia}

\newcommand{\SETI}{SETI Institute, Mountain View, California}
\newcommand{\KZA}{University of Malta, Institute of Space Sciences and Astronomy}
\newcommand{\PWJD}{The Breakthrough Initiatives, NASA Research Park, Bld. 18, Moffett Field, CA, 94035, USA}

\newcommand{\newtext}[1]{\color{black} #1} 


\correspondingauthor{Bryan Brzycki}
\email{bbrzycki@berkeley.edu}

\author[0000-0002-7461-107X]{Bryan Brzycki}
\affiliation{\UCB}

\author[0000-0003-2828-7720]{Andrew P.\ V.\ Siemion}
\affiliation{\UCB}
\affiliation{\SETI}
\affiliation{\NIJ}
\affiliation{\KZA}

\author[0000-0003-4823-129X]{Steve Croft}
\affiliation{\UCB}
\affiliation{\SETI}

\author[0000-0002-8071-6011]{Daniel Czech}
\affiliation{\UCB}

\author[0000-0003-3197-2294]{David DeBoer}
\affiliation{\UCB}

\author{Julia DeMarines}
\affiliation{\UCB}

\author{Jamie Drew}
\affiliation{\PWJD}

\author[0000-0002-8604-106X]{Vishal Gajjar}
\affiliation{\UCB}

\author[0000-0002-0531-1073]{Howard Isaacson}
\affiliation{\UCB}
\affiliation{\USQ}

\author{Brian Lacki}
\affiliation{Breakthrough Listen, \UCB}

\author{Matthew Lebofsky}
\affiliation{\UCB}

\author{David H.\ E.\ MacMahon}
\affiliation{\UCB}

\author{Imke de Pater}
\affiliation{\UCB}

\author[0000-0003-2783-1608]{Danny C.\ Price}
\affiliation{\UCB}
\affiliation{\SWIN}

\author{S. Pete Worden}
\affiliation{\PWJD}

\begin{abstract}
As it stands today, the search for extraterrestrial intelligence (SETI) is highly dependent on our ability to detect interesting candidate signals, or technosignatures, in radio telescope observations and distinguish these from human radio frequency interference (RFI). Current signal search pipelines look for signals in spectrograms of intensity as a function of time and frequency (which can be thought of as images), but tend to do poorly in identifying multiple signals in a single data frame. This is especially apparent when there are dim signals in the same frame as bright, high signal-to-noise ratio (SNR) signals. In this work, we approach this problem using convolutional neural networks (CNN) as a \newtext{computationally efficient method} for localizing signals in synthetic observations resembling data collected by Breakthrough Listen using the Green Bank Telescope. We generate two synthetic datasets, the first with exactly one signal at various SNR levels and the second with exactly two signals, one of which represents RFI. We find that a residual CNN with strided convolutions and using multiple image normalizations as input outperforms a more basic CNN with max pooling trained on inputs with only one normalization. Training each model on a smaller subset of the training data at higher SNR levels results in a significant increase in model performance, reducing root mean square errors by at least a factor of 3 at an SNR of 25 dB. Although each model produces outliers with significant error, these results demonstrate that using CNNs to analyze signal location is promising, especially in image frames that are crowded with multiple signals.
\end{abstract}



\section{Introduction}

Many avenues in the search for extraterrestrial intelligence (SETI) are largely reliant on our ability to pick out interesting signals in a sea of optical and radio telescope data. Since the 1960s, radio searches for evidence of extraterrestrial intelligence (ETI) have increased in scope in tandem with our improving technology, covering larger instantaneous bandwidths and surveying more targets than before \citep{Drake:1961, werthimer1985serendip, Horowitz:1986, Korpela:2001, welch2009allen, Siemion:2013, wright2014near, MacMahon:2018, Price:2018}. 

The Breakthrough Listen (BL) initiative is the most thorough SETI search effort, with access to top radio telescopes across the world specifically for use in SETI searches, including 20\% of the telescope time on the Green Bank Telescope (GBT) in West Virginia, USA and 25\% time on the CSIRO Parkes radio telescope in New South Wales, Australia \citep{Worden:2017, Isaacson:2017, MacMahon:2018, Price:2018}. In optical wavelengths, the search uses the Automated Planet Finder at the Lick Observatory in California, USA \citep{Vogt2014}. \newtext{The BL search has expanded to include such facilities as the MeerKAT telescope in South Africa \citep{jonas2009meerkat}, the VERITAS Cherenkov Telescope at the Whipple Observatory in Arizona, USA \citep{weekes2002veritas}, the Murchison Widefield Array in Western Australia \citep{tingay2018search}, and the FAST telescope in Guizhou Province, China \citep{zhang2020first}.} Sifting through the sheer data volume collected, which can be on the order of hundreds of terabytes per day, is computationally expensive alone, but identifying interesting, anomalous signals is itself a tough open problem.

Most of the coherent radio signals that we observe in BL data are anthropogenic, termed radio frequency interference (RFI). Types of RFI include satellite telemetry, cellular mobile broadcasts, wireless internet, and a host of other artificial sources. These are all types of narrow-band signals, which means each signal has a small frequency bandwidth (generally of order less than 1 kHz). On the other hand, natural astrophysical phenomena usually produce broad-band signals. The challenge for technosignature searches is that if an intelligent civilization is producing signals at radio frequencies (technosignatures), either as directed transmissions or as by-products of advanced technology, these signals are also likely to be narrow-band and therefore appear similar to RFI. SETI searches to date have found mountains of RFI signals, but no conclusive evidence of technosignatures \citep{tarter2001search, korpela2011status, siemion20131, siemion2014searching, harp2016seti, Enriquez:2017, gray2017vla, tingay2018search, wright2018much, Price:2019}.

The science data we collect from radio telescopes are generally stored as arrays of detected intensity (Stokes-I) as a function of time and frequency. These can be visualized as dynamic spectra or “waterfall plots,” with frequency on the x-axis, time on y-axis, and intensity as a color according to a colorscale. In other words, each pixel corresponds to an intensity value computed at that specific frequency and time. Narrow-band signals that are “on” for the duration of a short observation appear as lines across waterfall plots, which may be sloped due to the relative motion between the celestial source and the telescope, the so-called Doppler acceleration \citep{Sheikh:2019}. If a signal is bright enough, it is easily distinguishable by the human eye. However, it is simply impossible to visually inspect all the data we collect, which easily spans billions of frequency channels \citep{Lebofsky:2019}.

Our standard narrow-band signal search method uses TurboSETI\footnote{\url{https://github.com/UCBerkeleySETI/turbo_seti}}, an implementation of the ``tree deDoppler'' algorithm, which effectively averages along potential Doppler drift rates (slopes) in a spectrogram and searches for statistically high spikes in the resulting spectra \citep{Taylor:1974, Siemion:2013, Enriquez:2017, enriquez2019turboseti}. If one picks the correct drift rate and there is a signal at that rate, one should get a detection, since averaging reduces the impact of random noise and preserves the signal. While the underlying tree-based algorithm is more efficient than a naive search over all drift rates, this approach requires many passes over the data and potentially misses fainter signals masked by bright RFI \citep{pinchuk2019search}. 

\newtext{A complementary parametric algorithm for localizing narrow-band signals is the Hough transform, an edge-detection technique that translates an image into another 2D representation whose features correspond to edges in the original image \citep{hough1959machine, barinova2012detection}. Applying this transform to Stokes-I data and identifying bright features allows for the detection and localization of narrow-band signals, which manifest as ``edges'' in the data \citep{monari2006generalized, fridman2011seti}. This method also requires many passes over the data, however, and the features must still be extracted from the resulting transform (e.g. via thresholding).}

\newtext{Another method for detecting signals in radio data is to analyze the degree to which the data differs from an ideal statistical distribution, assuming only noise is present. For instance, higher order statistics such as the kurtosis can indicate that a portion of data differs significantly from an ideal Gaussian distribution. Applying this principle by calculating the kurtosis for time series voltage data or the spectral kurtosis for dynamic spectra can signal the presence of RFI \citep{ruf2006rfi, nita2016eovsa}. Since these are relatively simple calculations, they can be done in real-time to flag or even mitigate RFI during observations. }

\newtext{While these approaches each have their own strengths,} we would like to evaluate the effectiveness of machine learning (ML) methods in accurately identifying narrow-band signals, especially in the presence of bright RFI. Having a good signal localization and detection pipeline is crucial for identifying signals that are currently overlooked using conventional signal processing methods. 

Advances in computer vision techniques, especially with convolutional neural networks (CNN), have proven quite effective in classification and object detection tasks \citep{krizhevsky2012imagenet, simonyan2014very, szegedy2015going, ren2015faster, he2016deep}. \newtext{For radio signal processing, CNNs have been used to classify radio transmissions based on their modulation schemes, using time series voltage data \citep{o2018over}. In fact, Stokes-I data produced from radio observations is similar in structure to an image, so the data lends itself readily to computer vision techniques.}

\newtext{For instance, \cite{zhang2018fast} used CNNs on Stokes-I data to detect new pulses from the fast radio burst FRB 121102. They created a dataset of synthetic FRB pulses and trained a classifier to detect whether or not a data frame contained a pulse. Using this model on real observations, they were able to find 72 new pulses within a 5 hour radio observation. }

\newtext{Similarly, \cite{harp2019machine} created a synthetic dataset of radio spectrograms to resemble data taken by the Allen Telescope Array, inserting 6 different classes of narrow-band signals into an artificial noise background. They compared the performance of various deep CNN architectures in classifying the synthetic data frames, and found that their ML classifiers performed well for signals with relatively high signal-to-noise ratios (SNR). }

Modern object detection methods such as You Only Look Once \citep[YOLO;][]{Redmon:2016} use clever ways to quickly determine an arbitrary number of object bounding boxes in images. Even so, object detection and localization of long, thin objects remain particularly difficult. It is hard to draw meaningful bounding boxes around them, since such objects generally comprise only a small portion of bounding box areas, making it impractical to maximize the intersection over union measure with ground truth. In addition, since many radio signals can intersect at any position, it is harder to similarly split up an image frame into a coarse grid and only associate one signal with each grid cell, as in YOLO. This makes it especially difficult to detect an arbitrary number of signals in a frame. For this reason, we limit our present work to signal localization, in which we attempt to precisely predict the positions of a known number of signals in each image frame.

In this work, we investigate the effectiveness of machine learning signal localization on synthetic radio spectrogram data. We run experiments using CNN architectures and evaluate performance based on the root mean square error between true and predicted pixel locations as a function of signal intensity or SNR. \newtext{We further compare these localization results with signal detections from TurboSETI.} We conclude with future directions for improving signal localization and ultimately moving towards true object detection.

\section{Data and Preprocessing}
\label{sec:dataprep}

The SETI goal of looking for interesting signals in observations makes it difficult to get a large labeled dataset. To that end, it is an open question as to what sort of labels make the most sense -- there are so many different forms and patterns in human RFI that results would be highly dependent on the number and nature of classes. Furthermore, manual inspection can be ineffective in identifying lower intensity signals (whereas averaging along various drift rates can increase the SNR and thus reveal dimmer signals).

To test the sensitivity and accuracy of signal search procedures, we generated a set of synthetic observations that resemble real data from the GBT. In general, the Breakthrough Listen instrument at the GBT takes data over a range of frequencies (over a large bandwidth of a few GHz) at the same time \citep{MacMahon:2018}. Here, we focus on scientific data products that have a 1.4 Hz spectral resolution and a 1.4 second temporal resolution, at a frequency range of 4 -- 8 GHz (C-band). 

For this work, we analyzed image frames that are $32\times1024$ pixels -- 32 time samples tall and 1024 frequency samples wide. This effectively spans a total range of about $32\cdot 1.4\ \mathrm{s}\approx45$ s and $1024\cdot1.4\ \mathrm{Hz}\approx1430$ Hz. Although our observations easily span billions of frequency channels, for practical reasons, we limit the number of frequency channels per frame to better facilitate the use of CNNs.

\subsection{Noise Properties}
\label{subsec:noise}

\begin{figure*}
\begin{center}
  \includegraphics[width=\textwidth]{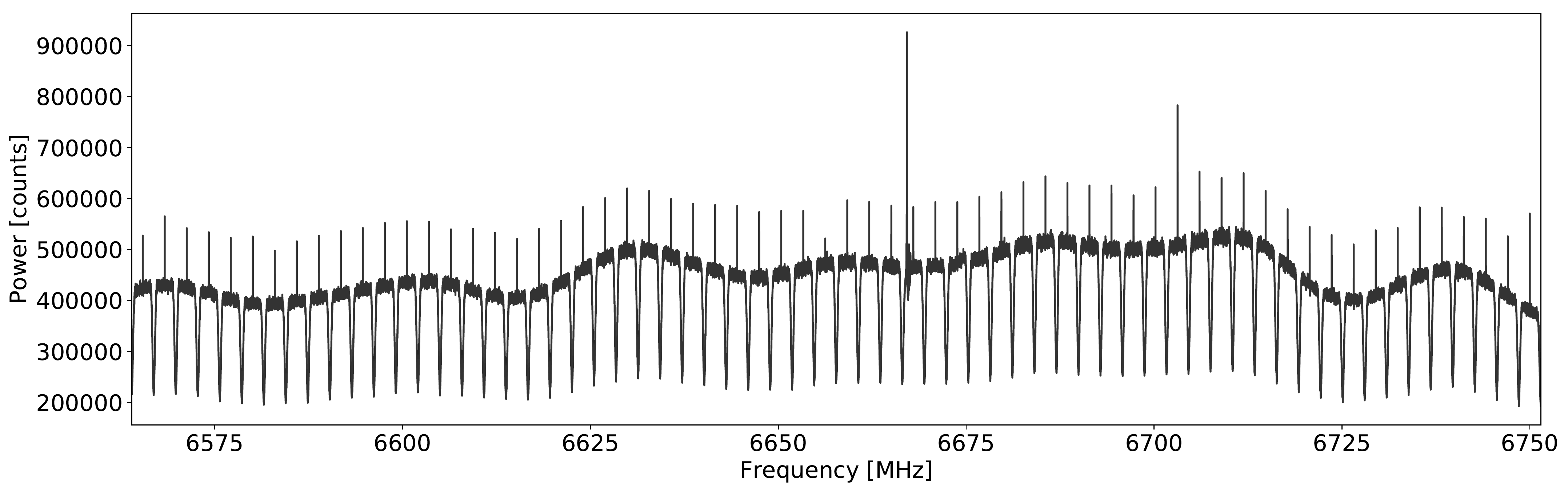}
  \caption{Bandpass plot for Sgr B2 data over an integration time of 60 s.}
  \label{fig:bandpass}
\end{center}
\end{figure*}

\begin{table}
\caption{Parameters for Sgr B2 data} 
\begin{center} 
\begin{tabular}{ c c } 
    \hline 
    Parameter & Sgr B2 \\ \hline \hline
    RA (J2000) & 17$^h$ 47$^m$ 15.0$^s$ \\ 
    Dec (J2000) & -28$^\circ$ 22' 59.16" \\ 
    Initial MJD & 58465.71709 \\ 
    C-Band Frequency Coverage & 6564 -- 6752 MHz \\ 
    Frequency Resolution & 1.39698 Hz \\
    Time Resolution & 1.43166 s \\
    Integration Time & 60 s \\
    \hline
    
  \end{tabular} 
  \label{table:params}
\end{center}
\end{table}

\begin{figure}
\begin{center}
  \includegraphics[width=.45\textwidth]{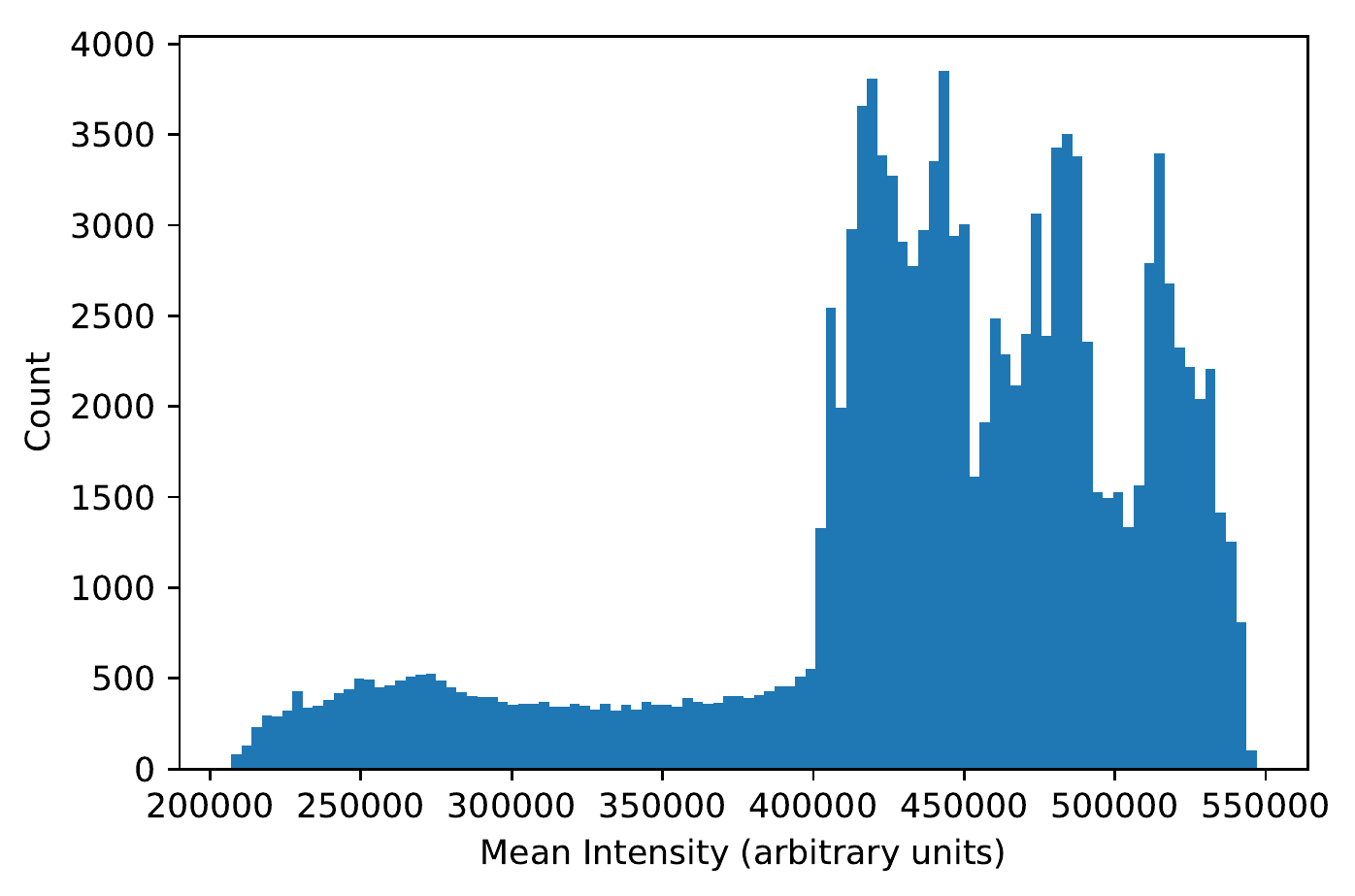}
  \caption{Histogram of mean frame intensities over real GBT observation after trimming outliers, for a total of 126,419 samples.}
  \label{fig:all_means}
\end{center}
\end{figure}

\newtext{Since the background radiometer noise in time-series voltage data closely follows a zero-mean Gaussian distribution, the noise in Stokes-I data ultimately follows a chi-squared distribution \citep{mcdonough1995detection, nita2007radio, ThompsonMoranSwenson}. However, due to instrumental effects such as coarse channel bandpass shapes and natural variations in detector sensitivity as a function of frequency, the raw intensity values we get from observations can vary appreciably.} 

To properly capture these intrinsic intensity variations in our synthetic data, we based their noise properties on actual data from the GBT. We used 4 -- 8 GHz observations of Sgr B2 taken on Dec 13, 2018 at 17:12:37 UTC. We found that using a smaller range of frequencies, 6564 -- 6752 MHz, was sufficient for obtaining realistic background intensity values for the synthetic frames (Table \ref{table:params}). 

At the GBT, the Versatile Green Bank Spectrometer \citep[VEGAS;][]{Prestage:2015} digitizes and coarsely channelizes data using a polyphase filterbank. The data is then sent to the Breakthrough Listen data recorder system \citep{MacMahon:2018}, which applies finer channelization to each coarse channel and records the resulting high spectral resolution data products. 

Figure \ref{fig:bandpass} shows the integrated bandpass plot of our observational data. Visible in the spectrum are the 64 coarse channels present in our data slice, which are about 3 MHz in width and characterized by intensity fall-offs on either edge. The spike at the center of each coarse channel is the so-called ``DC bin,'' the sum of all samples within the channel, which arises from the Fourier Transform-based filterbank. The large spike at $\sim$6670 MHz is bright RFI. The overall bandpass shape reflects the inherent variation in sensitivity across the receiver. 

\newtext{We split this data into frames of size $32\times1024$ and calculate the mean intensity of each individual frame. Since some frames contain DC bins and bright RFI that bias these intensities toward higher values, we trim the resulting collection of mean intensities using sigma clipping with limits corresponding to $5\sigma$.} Figure \ref{fig:all_means} shows a histogram of these intensities after cutting out outliers, for a total of 126,419 remaining frames. Together with Figure \ref{fig:bandpass}, we notice that the majority of mean intensities are concentrated between $4\times 10^5$ and $5.5\times 10^5$. The tail extending down to intensities of $2\times 10^5$ is due to the lower sensitivities at the edges of each coarse channel. Note that each frame only covers $\sim$1.4 kHz in frequency, which is small in comparison to coarse channels ($\sim$3 MHz). So, we make the assumption that larger scale systematic bandpass effects are not important within individual frames.

\newtext{To initialize each synthetic data frame with noise, we randomly select from this empirical distribution to select a desired mean intensity $\mu_\textrm{noise}$. To calculate the degrees of freedom for our chi-squared noise background, we note that our Stokes-I data uses two polarizations and that complex Fourier coefficients contribute both real and imaginary terms. These result in 4 degrees of freedom for every integration, so overall, the underlying distribution has a total of $k=4\cdot df\cdot dt$ degrees of freedom, where $df$ and $dt$ are the frequency and time resolutions of the data. Since the mean of a chi-squared distribution is $k$, we randomly sample values from this distribution to populate the empty data frame, and then scale every value up by a factor of $\mu_\textrm{noise}/k$ to match the desired mean intensity. Comparing the results to our observations, we find that this procedure indeed reproduces the noise distributions found in real data. The benefit in having a method for generating purely synthetic yet realistic background noise is that every data frame thus created is guaranteed to be free from signals of any kind, which helps in accurately evaluating signal search strategies. }

\subsection{Synthetic Signals}
\label{subsec:signals}

\newtext{Narrow-band signals found in our radio frequency data come in many forms, with temporal and spectral structures of varying complexity. For example, modulation schemes in prolonged radio transmissions result in intensity variations over time, and some signals are emitted in short pulses in the first place \citep{sokolowski2015statistics}. Depending on their location in the galaxy and in the sky, narrow-band signals may also be subject to scintillation from the interstellar and interplanetary medium \citep{rickett1977interstellar, lotova1985interplanetary, cordes1997scintillation, Siemion:2013, 2019MNRAS.486.3636P}. Even a constant amplitude sine wave signal in the time domain, in the presence of zero-mean Gaussian noise, follows a non-central chi-squared distribution of intensities in Stokes-I data \citep{mcdonough1995detection}. Examples of various RFI morphologies are presented in \cite{sheikh2020breakthrough}. }

\newtext{Despite the prevalence of complex and noisy narrow-band signals, for this work, we choose to create synthetic signals with constant intensity over time as heuristic models for real signals. Indeed, this makes the assumption that if an ML model can accurately localize these ideal signals, it will also properly localize more complex signals. Our intuition behind this assumption stems from the fact that search techniques such as TurboSETI will still find noisy signals, even though they are optimized for finding ``simple'' signals. }

We developed a software package, Setigen\footnote{\url{https://github.com/bbrzycki/setigen}}, to facilitate the creation and injection of synthetic narrow-band signals into observational data frames. \newtext{Based on the time and frequency resolutions of the Stokes-I data, Setigen can also calculate the corresponding idealized chi-squared distribution for the background noise, into which synthetic signals can be added, as described in Section \ref{subsec:noise}.} This allows us to create large datasets of synthetic data frames for training and validating signal search pipelines.

We define the ``start'' of a signal as the index (or pixel) in the frequency direction where the center of the signal is during $t=0$ in an image frame, and the ``end'' as the center position during the 32nd time sample. We randomly choose the starting and ending indices for each signal and the width of the signal in the frequency direction (limited to a narrow-band range). Starting and ending indices are always between 0 to 1023, inclusive. Accordingly, the maximum absolute drift rate, corresponding to starting and ending indices on opposite ends of the frequency range, would be about 31 Hz/s. 

Because we would like to analyze the effectiveness of our machine learning algorithms on different SNR levels, we scale the intensity of each synthetic signal according to the desired SNR level, the mean background noise level, and the number of time samples:

\begin{equation}
    \textrm{SNR} = \frac{I_\textrm{signal}}{\mu_\textrm{noise}}\times\sqrt{n_t},
\end{equation}

where $I_\textrm{signal}$ is the appropriate intensity of the injected constant signal at any single time samples, $\mu_\textrm{noise}$ is the mean of the background noise, and $n_t$ is the number of time samples (in this case, $n_t=32$).

This definition is used so that the expected SNR matches the measured SNR if we had simply averaged through each time sample shifted at the correct drift rate, which is how current Doppler drift search pipelines, such as TurboSETI, work \citep{Enriquez:2017}.

\subsection{Dataset Construction}

We generate two datasets to test signal localization, each with 120,000 training samples and 24,000 test samples. Since the signals we are interested in potentially span a large range of intensities, we specify SNR levels for our synthetic narrow-band signals using decibels, such that 0 dB $\rightarrow\ 1\sigma$, 20 dB $\rightarrow\ 100\sigma$, etc.

Our first dataset contains $32\times1024$ image frames with exactly one signal at SNR levels of 0, 5, 10, 15, 20, and 25 dB. So, for each SNR level, we generate 20,000 training frames and 4,000 test frames. For each frame, we also save the starting and ending indices (2 numbers) as labels.

Our second dataset contains frames with exactly two signals. One of the signals is 25 dB and at a zero drift rate, so that it is at a constant frequency at all time samples. This is meant to represent a typical RFI-like signal. The other signal is at SNR levels of 0 -- 25 dB as in the first dataset. We save the starting and ending indices for both signals (4 numbers). Note that these labels $\mathit{are}$ ordered, even though there is no preferred order in any given image frame. 

\begin{figure}
\begin{center}
  \includegraphics[width=.45\textwidth]{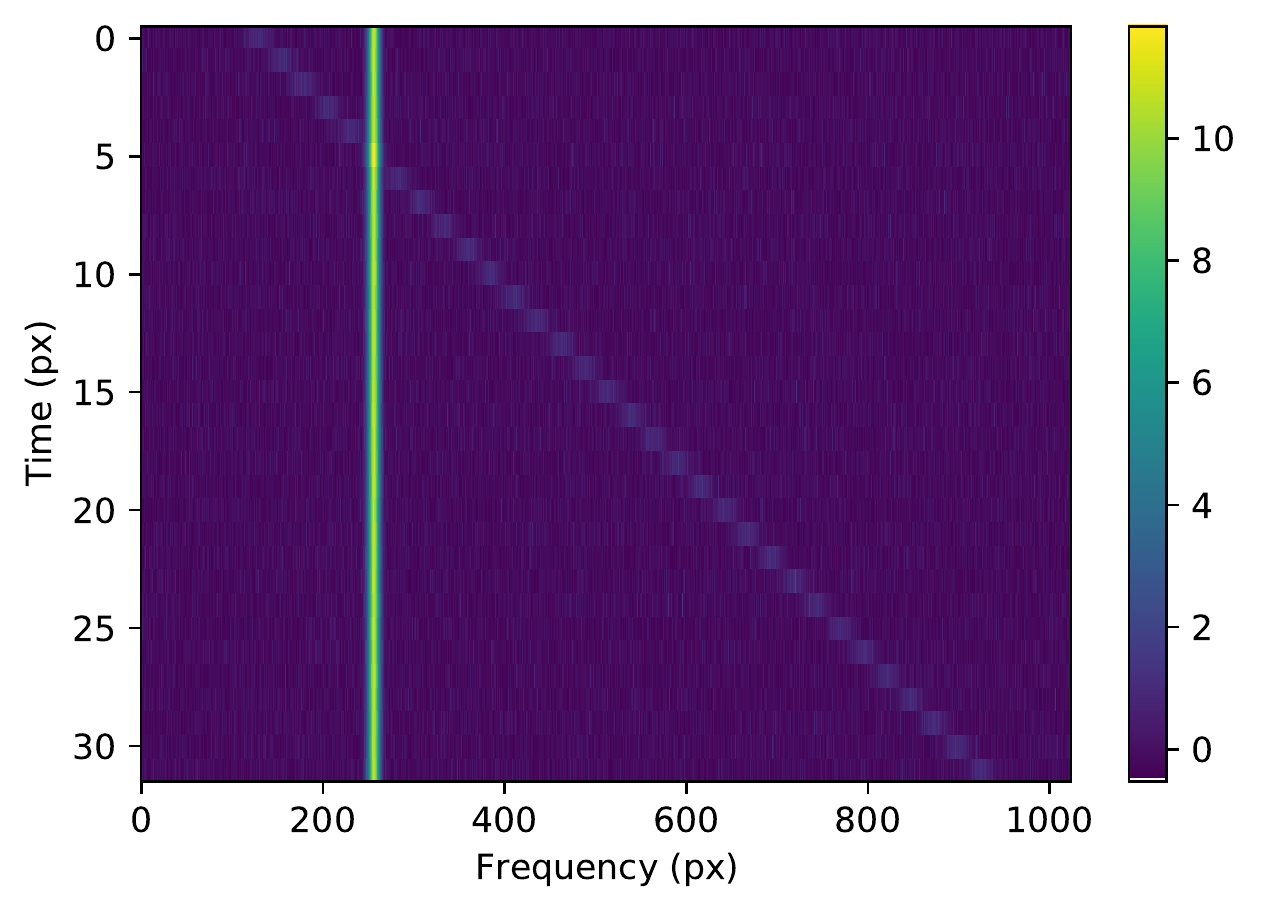}
  \caption{Synthetic data frame with two signals, one ``RFI'' signal at 25 dB and zero drift, and one dimmer signal at 15 dB, normalized over the entire frame to mean 0 and variance 1.}
  \label{fig:normalization1}
\end{center}
\end{figure}

\begin{figure}
\begin{center}
  \includegraphics[width=.45\textwidth]{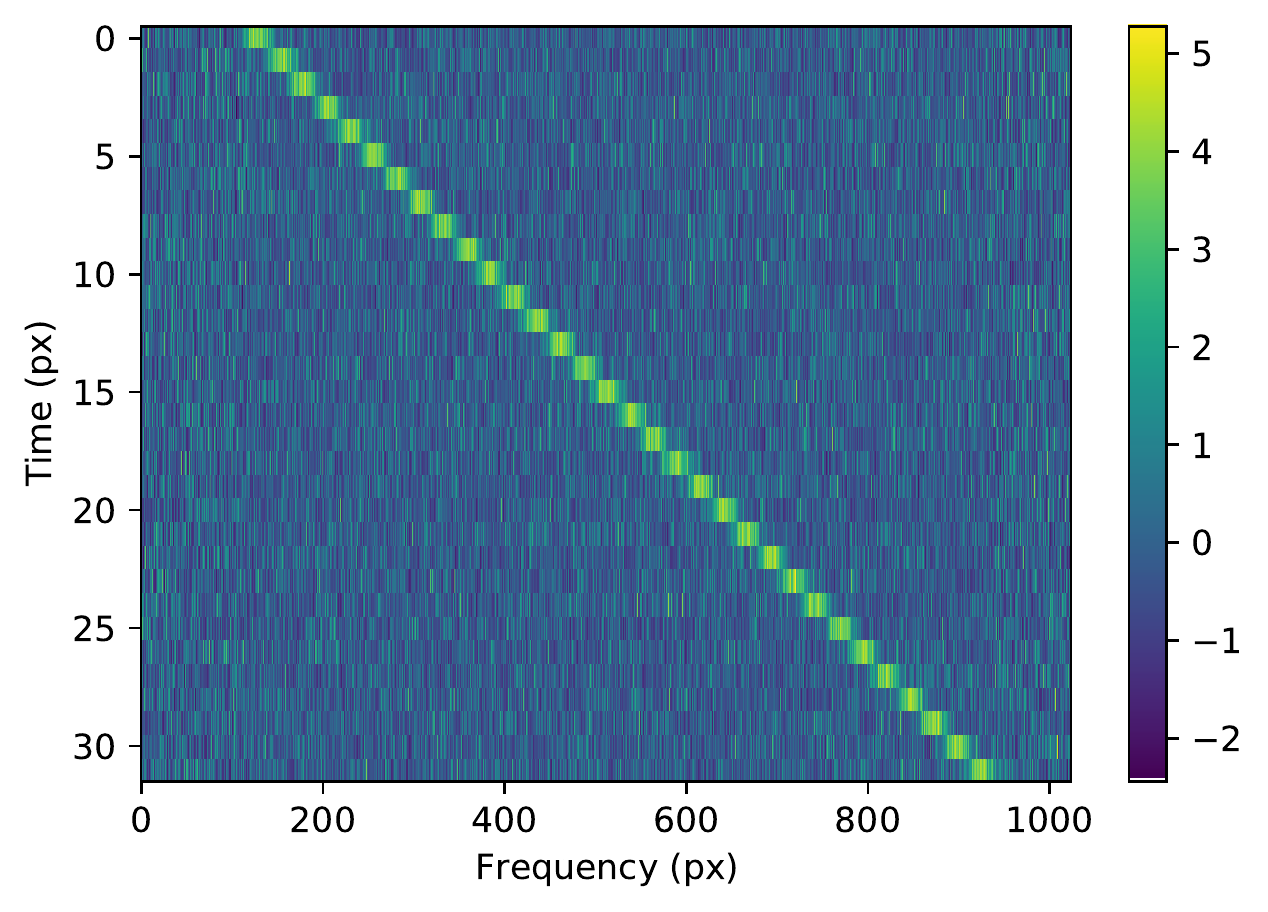}
  \caption{Data frame containing the same data as Figure \ref{fig:normalization1}, instead normalized per frequency channel to mean 0 and variance 1.}
  \label{fig:normalization2}
\end{center}
\end{figure}

\subsection{Preprocessing}

By design, for each dataset, we generate 120,000 training frames and 24,000 test frames in total, the latter of which are used only for final evaluation. During training, we take a 80/20 random split of the training frames for training/validation.

The choices of normalization are very important for our data, which can exhibit regions of high contrast and varying instrument sensitivity. We choose to normalize our labels (indices between 0 and 1023 inclusive) to be between 0 and 1 by dividing out by 1024. Normalizing our input data frames is more interesting, and there are multiple potential ways to go about this.

The first would be to normalize over an entire frame by subtracting the mean and dividing by the variance over all pixels, so that our normalized frame has mean 0 and variance 1, as in Figure \ref{fig:normalization1}. Another method useful in astronomy is normalizing by frequency, where we subtract the mean and divide by the variance in the time direction for each frequency sample. This also yields mean 0 and variance 1, but serves to specifically normalize out differences in instrument sensitivity as a function of frequency. Normalizing over an entire frame preserves these sensitivity differences.

However, lots of detected narrow-band signals are RFI and thus moving with the Earth, so they do not exhibit Doppler accelerations and appear as vertical lines in waterfall plots. So if we normalize by frequency, a constant vertical signal will disappear from the data, since we subtract out the average (constant) intensity, as in Figure \ref{fig:normalization2}. Since we are interested in localizing all signals and eventually comparing our machine learning methodology with standard search techniques, we certainly do not want to exclude this information via our normalization procedure. On the other hand, this can potentially strengthen our sensitivity towards dimmer, sloped signals. 

Considering these idiosyncrasies in our data, we test both of these two normalization methods as inputs into our models.

\section{Methods}

\newtext{We present the signal search methods used in this work. Namely, we discuss the CNN architectures explored for ML-based localization and briefly describe our standard deDoppler algorithm, TurboSETI.}

\subsection{CNN Model Architectures}
\label{sec:models}

In this work, we define both a ``baseline" and a ``final" model. We take our baseline architecture to be a simple CNN that is typically used on general image classification tasks. Our final model contains architectural improvements influenced by the nature of our data and training tasks. Specifically, we compare the baseline and final models and evaluate the extent to which the architectural changes improve localization accuracy.

For our two datasets, we use the same overall model architecture except for the final regression layer, which either has 2 or 4 nodes depending on whether we are predicting the position(s) of one or two signals. Since we would like to predict each signal position as best as possible, where the position is defined by its starting and ending indices, we seek to minimize the mean squared error between true and predicted indices. 

Our models are implemented using the Keras functional API \citep{chollet2015keras}. The source code for generating our datasets and training these models is available at \href{https://github.com/bbrzycki/seti-nb-localization}{github.com/bbrzycki/seti-nb-localization}.

\subsubsection{Baseline Model}

\begin{figure}
\begin{center}
  \includegraphics[width=.45\textwidth]{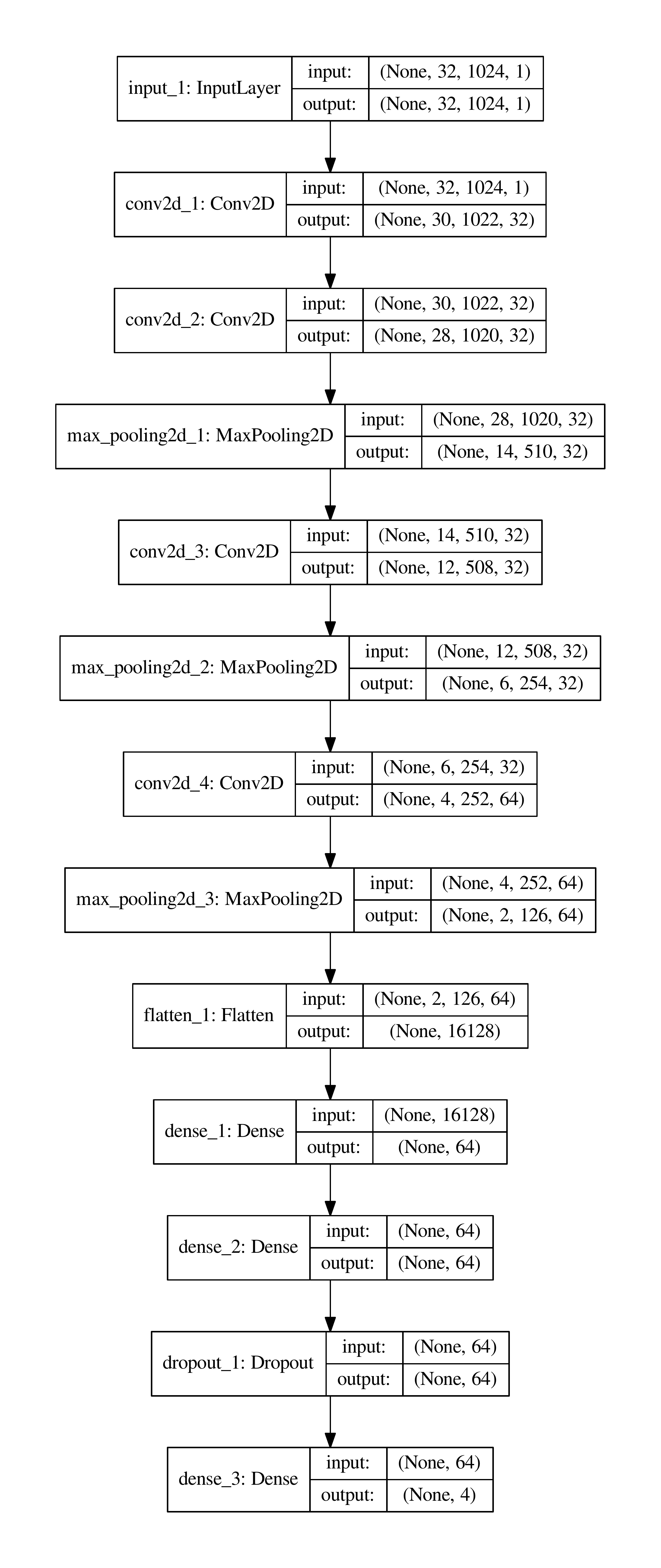}
  \caption{Baseline model architecture for the two signal localization task. For each block, the layer type is shown, along with input and output shapes. Inputs have shape $32\times 1024\times 1$, normalized over all pixels. These are passed through an initial convolutional layer, followed by 3 pairs of convolutional and max pooling layers. This is followed by two fully connected layers and a dropout layer, before finally going into the output layer. For the single signal task, the last layer has 2 nodes instead of 4.}
  \label{fig:baseline_shapes}
\end{center}
\end{figure}

For our baseline model, we choose a simple CNN with 4 convolutional layers, 3 max pooling layers, 2 fully connected layers (each with 64 nodes), and a dropout layer at 50\%. Our input is a single $32\times1024$ data frame, normalized over the entire frame. This is a typical architecture for image classification tasks, making it a good baseline for comparison. We use rectified linear unit (ReLU) activations after each convolutional and fully connected layer. For the two signal task, this model has 1,073,988 trainable parameters. Figure \ref{fig:baseline_shapes} shows the baseline model architecture in detail.

\subsubsection{Final Model}
\label{subsubsec:final}

For our final model, we have 2 residual connections, 5 convolutional layers in total (using stride 2 instead of max pooling), 2 fully connected layers (both with 1024 nodes), and a dropout layer at 50\%. We again use ReLU activations after each convolutional and fully connected layer, as well as batch normalization after summing frames in residual connections. Although our image frames were normalized to a mean of 0 and therefore contained negative numbers, we found that alternate activation functions to ReLU, such as tanh, did not improve localization accuracy. 

Residual connections are marked by shortcuts between convolutional layers; in our case, we use element-wise addition between a given layer and a following convolutional layer \citep{he2016deep}. These connections reduce over-fitting and enhance accuracy by counteracting vanishing gradients in neural networks. Furthermore, since narrow-band signals are thin relative to our image frames, residual connections allow thinner features to propagate further into our models. We follow up these additive connections with batch normalization layers to ensure that the lower-order statistics of layer inputs at these positions in our model remain the same across batches of data \citep{ioffe2015batch}.

For our inputs, we express the data frames as ``images'' with two channels -- one channel is the data frame normalized over the entire frame, and the other is the data frame normalized per frequency. Our rationale for using a two channel input is that when one normalizes over the entire frame, the model finds the brighter signals much more easily, and the dimmer signals could be washed out. However, most of time in radio data, the brightest signals are also at zero drift rate, since they originate from Earth. Therefore, normalizing by frequency could serve to remove these brightest signals and show more sensitivity to dimmer, drifted signals. Using both forms of image normalization as inputs into the same model could help better identify these different forms of signals appearing in our data. 

\begin{figure}
\begin{center}
  \includegraphics[width=.4\textwidth]{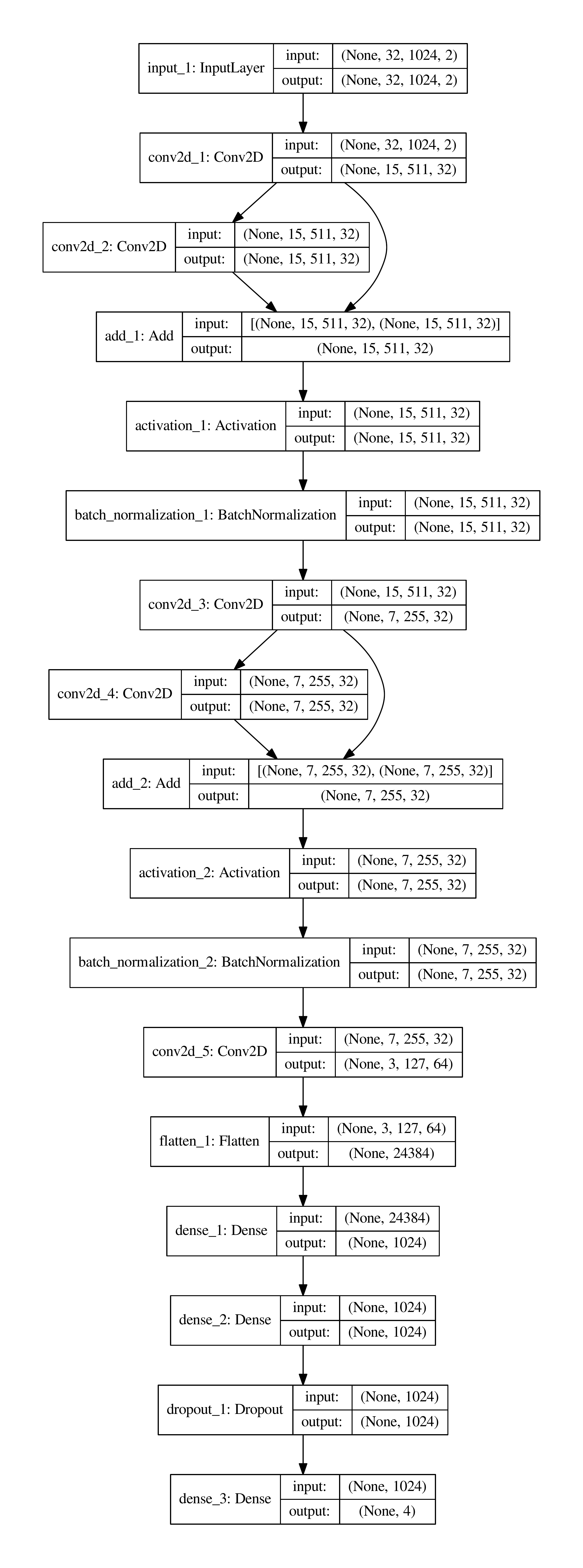}
  \caption{Final model architecture for the two signal localization task. Inputs have shape $32\times 1024\times 2$, combining the two normalizations described in Section \ref{subsubsec:final}. Residual connections are apparent between convolutional layers, followed by a batch normalization layer. This structure is repeated twice, followed by another convolution layer, two fully connected layers, and a dropout layer. For the single signal task, the output layer has 2 nodes instead of 4.}
  \label{fig:final_shapes}
\end{center}
\end{figure}

For the two signal task, this model has 26,070,916 trainable parameters. Figure \ref{fig:final_shapes} shows the final model architecture in detail.

\subsection{TurboSETI}

\newtext{To provide a standard for performance comparison, we also ran our one signal dataset through the TurboSETI suite. As mentioned before, TurboSETI uses a tree-based deDoppler algorithm to efficiently search for signals above a specified SNR threshold over a specified range of drift rates \citep{Enriquez:2017, enriquez2019turboseti}. For each detected signal, TurboSETI returns the starting index/frequency and drift rate, along with the calculated SNR. }

\newtext{Previous studies have used the detection threshold of an SNR of 10; going any lower results in an unacceptable number of false positive detections \citep{Price:2019, sheikh2020breakthrough}. We likewise set a detection threshold at an SNR of 10, or 10 dB. }

\newtext{Selecting a maximum Doppler drift rate range to search presents a trade-off between potential detections and computational time. \cite{sheikh2020breakthrough} uses a maximum absolute drift rate of 20 Hz/s, which is the largest thus far for a deDoppler search strategy. As mentioned in Section \ref{subsec:signals}, the largest possible absolute drift rate in our $32\times1024$ frames is 31 Hz/s, so we simply choose that as our maximum search drift rate. }

\newtext{Because TurboSETI works by integrating along straight line paths, it struggles to find dim signals that are close to or intersect brighter signals. We observe this with every frame in our two signal dataset, in which TurboSETI only finds the bright, non-drifted ``RFI'' signal. Accordingly, we only compare our TurboSETI search results with the ML predictions over the one signal dataset. }

\section{Results}
\label{sec:results}

\subsection{Baseline vs. Final Model}

\begin{figure}
\begin{center}
  \includegraphics[width=.45\textwidth]{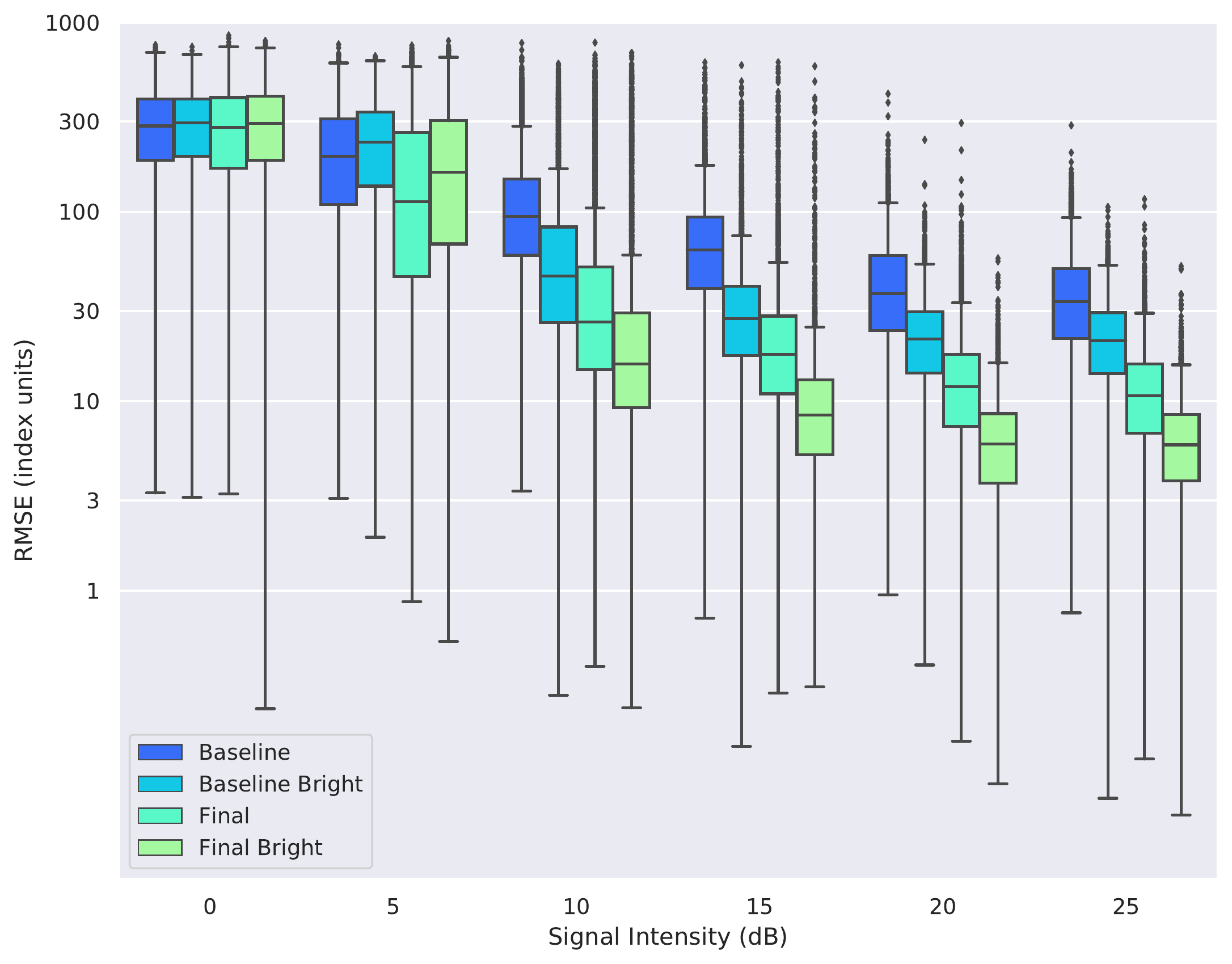}
  \caption{Box plot of RMSE in index/pixel units for the one signal dataset as a function of signal SNR. We compare the final and baseline models trained on both the full 0 -- 25 dB dataset, as well as the truncated ``bright'' 10 -- 25 dB dataset.}
  \label{fig:1sig_rmse}
\end{center}
\end{figure}

\begin{figure}
\begin{center}
  \includegraphics[width=.45\textwidth]{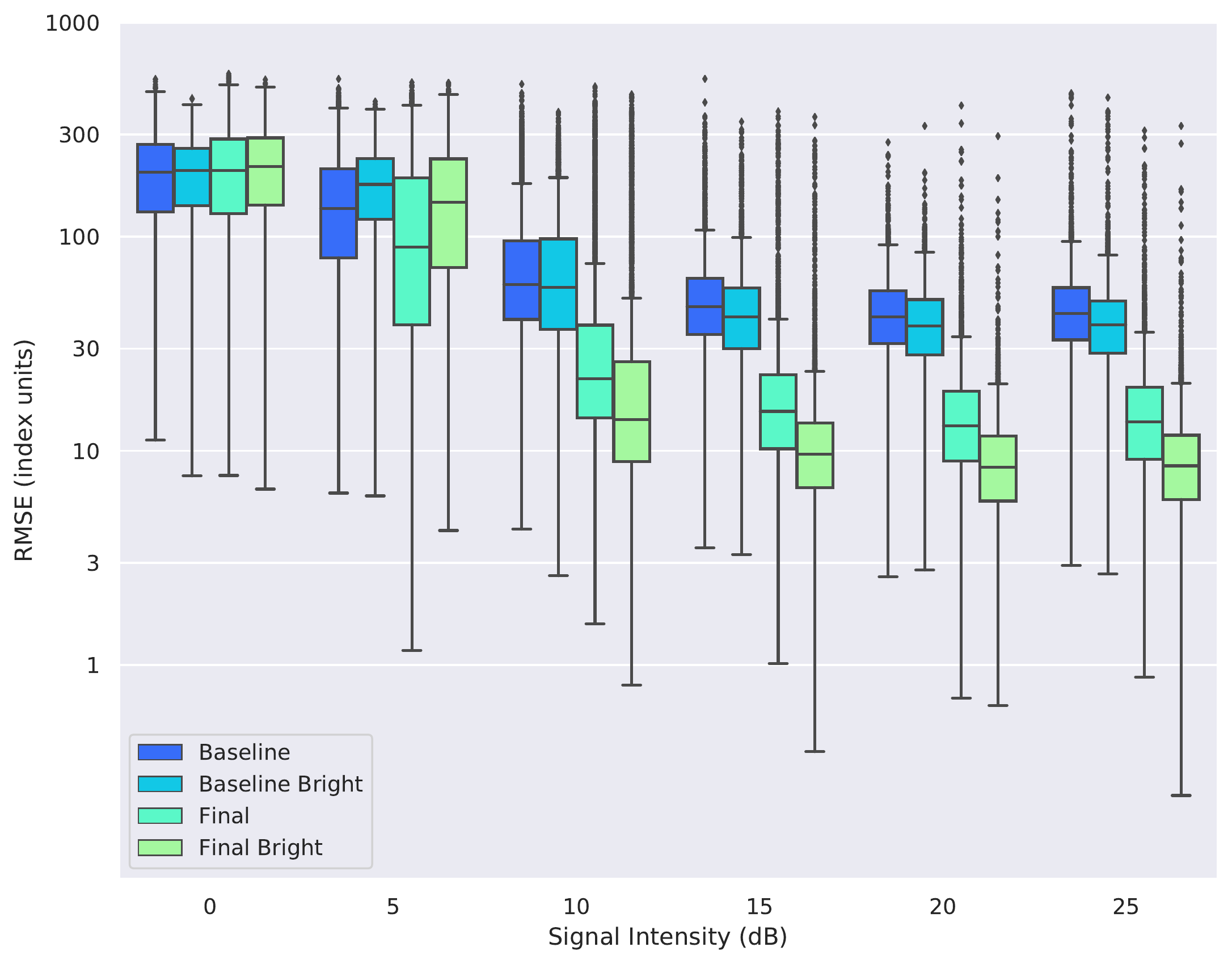}
  \caption{Box plot of RMSE in index/pixel units for the two signal dataset as a function of signal SNR. We compare the final and baseline models trained on both the full 0 -- 25 dB dataset, as well as the truncated ``bright'' 10 -- 25 dB dataset.}
  \label{fig:2sig_rmse}
\end{center}
\end{figure}

\begin{figure}
\begin{center}
  \includegraphics[width=.45\textwidth]{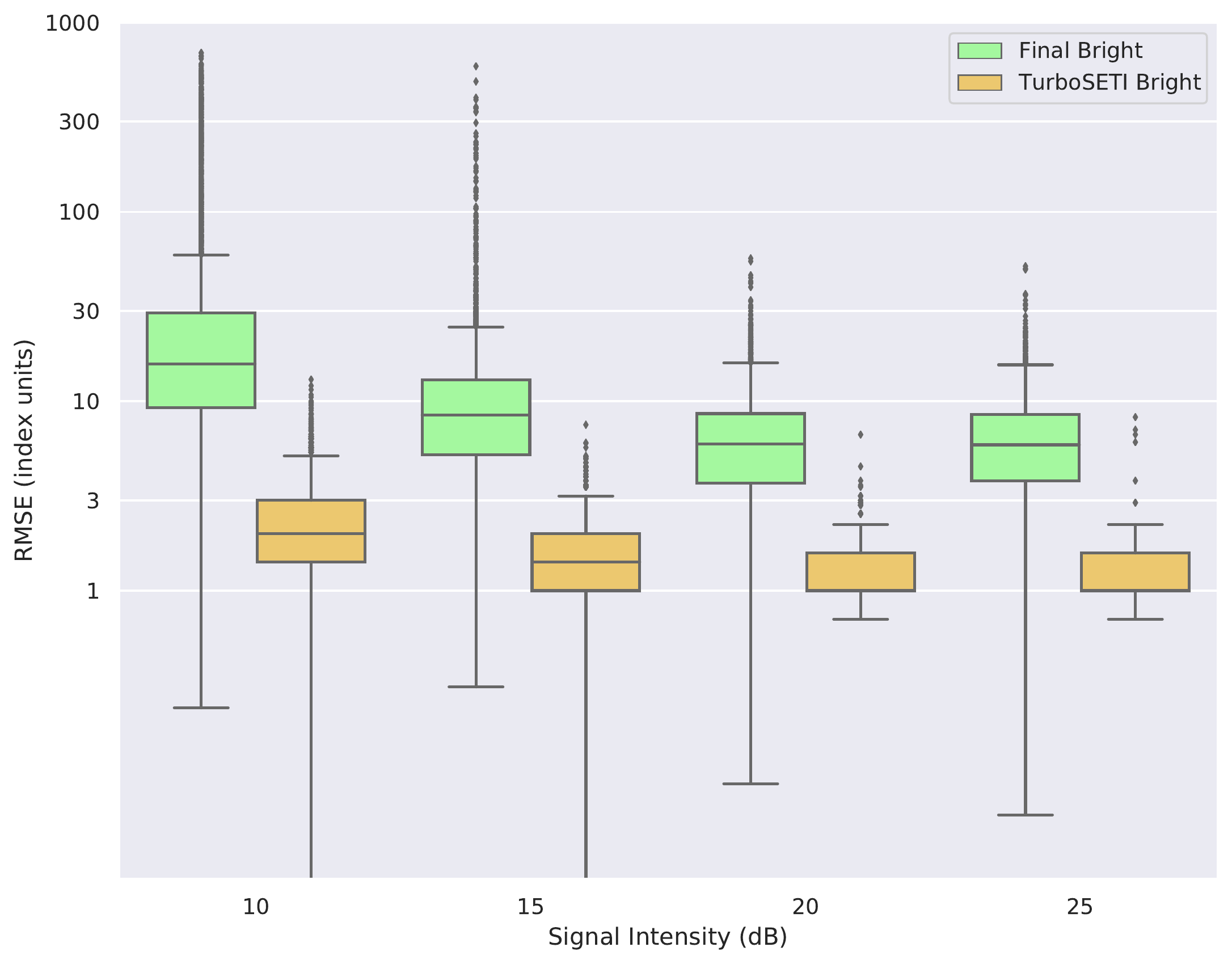}
  \caption{Box plot of RMSE in index/pixel units, comparing performance of our final model architecture with TurboSETI on the one signal dataset as a function of signal SNR.}
  \label{fig:turboseti_rmse}
\end{center}
\end{figure}

For the baseline and final model architectures, we compare the error between true and predicted indices as a function of SNR. To get a more intuitive feel on model performance, we calculate $1024\times$RMSE, where RMSE is the root mean square error, to see the errors in units of pixels/indices:

\begin{equation}
\mathrm{RMSE}\ \mathrm{(index}\ \mathrm{units)} = 1024\times\sqrt{\frac{1}{n}\sum_{i}^{n}(y_i - \hat{y_i})^2},
\end{equation}

where $n$ is the number of indices in our labeled data (2 and 4 for one and two signal datasets), $y_i$ are predicted indices, and $\hat{y_i}$ are true indices.

We first train our models on the full 0 -- 25 dB SNR levels at each dataset. We also analyze the effect of only training on frames with a 10 -- 25 dB signal, cutting out the 0 and 5 dB levels. Here, we again use a train-test split of 80/20 out of this restricted training set. However, we still evaluate these trained models on the full test set. Results from the restricted 10 -- 25 dB dataset are labeled ``bright'' in Figures \ref{fig:1sig_rmse}-\ref{fig:2sig_rmse}.

In these figures, we plot a box and whisker plot (with outliers above and below the median by 1.5 times the spread between 25\% and 75\% quartiles) for each SNR level and each training run for our baseline and final models on the full and bright datasets. For each case, we have outliers with high errors that would tend to bias our evaluations much higher if we only consider the mean RMSE.

\subsection{TurboSETI vs. Final Model}
\label{subsec:turbo}

\newtext{For each frame in the one signal dataset, we use TurboSETI to get signal localization results, which we translate into starting and ending indices. This allows us to calculate the RMSE in index units, exactly as we did for the ML predictions. Figure \ref{fig:turboseti_rmse} shows the results, again as a function of SNR level, compared to predictions from our final CNN model trained on ``bright'' frames. Note that we only compare predictions down to an intensity of 10 dB, since that is our TurboSETI detection threshold. }

\newtext{We also compare the computational costs of each search method. For the final CNN model, generating predictions for all 24,000 frames in the test set takes about 70 seconds using an Nvidia Titan Xp GPU with a batch size of 1. Using a batch size of 32, generating all predictions takes about 34 seconds.}

\newtext{Although we run TurboSETI with a 10 dB threshold, for bench-marking purposes, we ran it on all 24,000 frames in the test set. This takes a total of 2.6 hours using a single CPU. In practice, however, TurboSETI is run on large data frames, on order of billions of frequency channels, as opposed to the comparatively small frames used in this work. This is because the tree-based search algorithm is most efficient when applied to a few large files, as opposed to many smaller ones. To make a fairer comparison, we ran TurboSETI on a large frame with the same amount of data as if all 24,000 test frames were concatenated along the frequency axis. Using the same search parameters of a 10 dB threshold and a maximum 31 Hz/s drift rate, TurboSETI takes about 20 minutes to finish searching this concatenated data frame. Making predictions using the CNN model is therefore more efficient, especially when using larger batch sizes. }

\begin{figure}
\begin{center}
  \includegraphics[width=.45\textwidth]{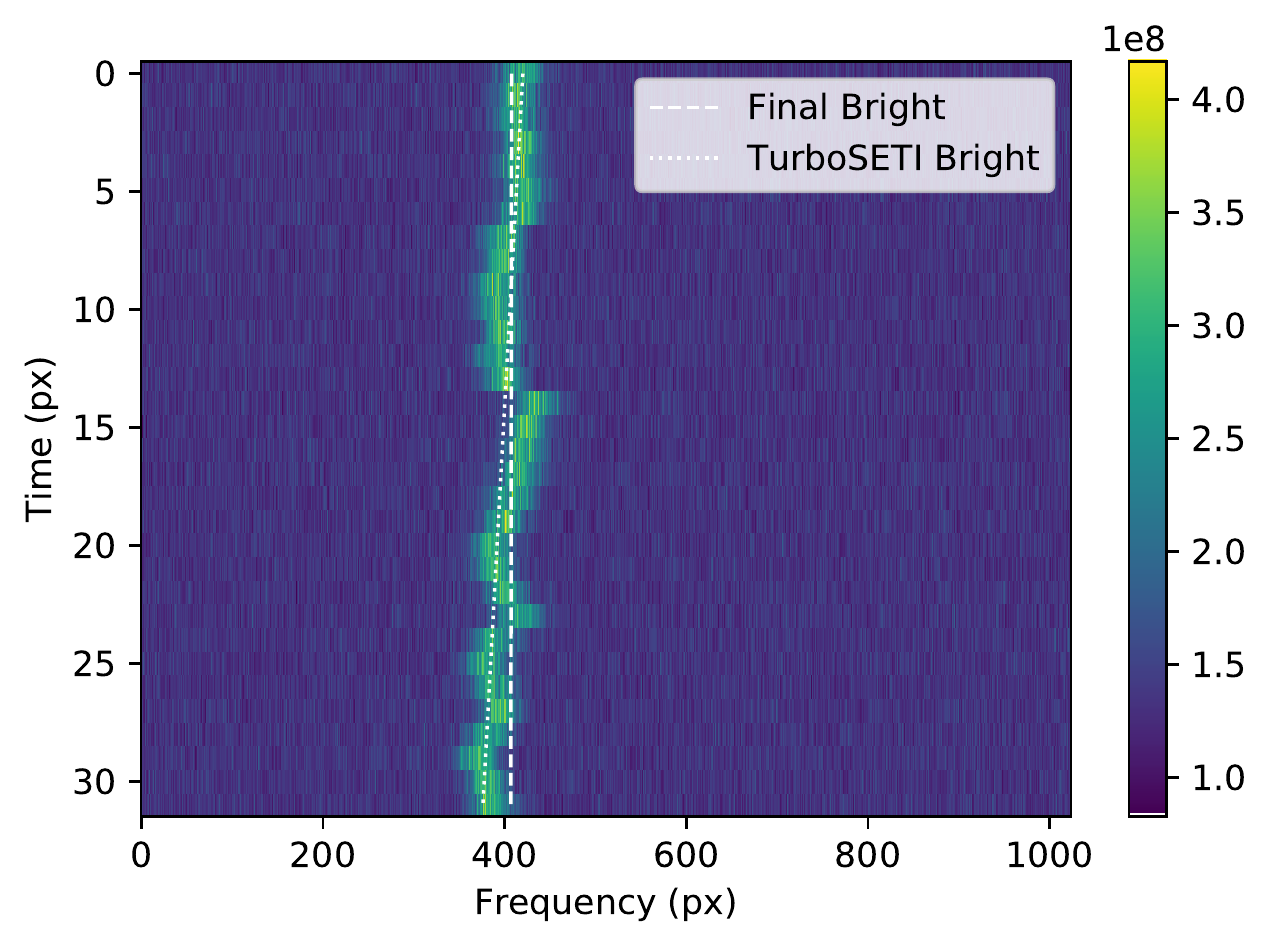}
  \caption{Observational data frame containing an RFI signal, overlaid with localization predictions from our final CNN architecture (dashed) and TurboSETI (dotted). Although the real signal is more complex than those in our training data, the model produces reasonable predictions.}
  \label{fig:rfi_ex}
\end{center}
\end{figure}

\newtext{Lastly, we tested both TurboSETI and the final CNN model on localizing a few known RFI signals in the C-band observation described in Section \ref{subsec:noise}. Figure \ref{fig:rfi_ex} shows an example of predicted localization paths from both methods for a complex RFI signal. As expected, TurboSETI generates the best-fit localization, since it integrates along each potential drift rate. Nevertheless, the final model gives a reasonable prediction, despite being trained on idealized signals with constant intensity and constant drift rate. }

\section{Discussion}
\label{sec:discussion}

\newtext{For the full training datasets, the baseline and final models show a smooth progression of better median RMSE values from 0 dB to 25 dB. Our final model seems to outperform the baseline model consistently, particular for signals of at least 10 dB, on the order of a $3\times$ reduction in error for both one and two signal datasets (Figures \ref{fig:1sig_rmse}-\ref{fig:2sig_rmse}).}

Unsurprisingly, for both one and two signal datasets, the signals at 0 dB (SNR=1; $1\sigma$) do very poorly, with an typical error of 200 -- 400 pixels off. As a check, note that the average expected distance between two randomly chosen points on a line segment of length $L$ is $L/3$. For 1024 total pixels, this means on average, selecting an index at random as a prediction should yield errors of about $1024/3\approx341$ pixels. So for 0 dB signals, our predictions are essentially random. In a way, this is perfectly acceptable, since in general we do not accept $1\sigma$ as a true detection of a signal in the first place. 

When we compare these results with those from models only trained on 10 -- 25 dB frames, we see a few interesting things. As expected, the ``bright'' models then perform worse on the lower SNR signals, 0 and 5 dB. For 0 dB signals, we still get effectively random predictions, but the decrease in performance at 5 dB is notable. \newtext{However, this restricted training set improves the performance at 10 -- 25 dB appreciably, in different ways. In the one signal case, the final model outperforms the baseline model by about $3\times$ for all SNR levels of at least 10 dB. In the two signal case, frames with at least 15 dB signals improved about $5\times$ between final and baseline models. We further observe that, in each case, restricting training to frames with at least a 10 dB signal improves overall model performance compared to training with the full dataset. Since both models seem to only be capable of nearly random predictions at the lowest SNRs, doing training steps on that data tends to hurt the performance at higher SNRs. }

The best performing models were the final model trained on the bright dataset. \newtext{At an SNR level of 25 dB, the one and two signal cases reached a median of about 6 and 9 pixels of RMSE, respectively.} We expect the one signal models to do better than the two signal models, since there is only a single bright signal to try to localize. Nevertheless, it was surprising that even our best models did not consistently localize either of these cases to extremely high precision, i.e. median RMSE of about 1 pixel. \newtext{On the one hand, localizing a bright signal to a precision of 6 pixels out of 1024 is decent (corresponding to about 8.4 Hz), but on the other, we expected to do even better, since we have synthetic datasets and therefore know $\mathit{precisely}$ where the center of each signal lies. Instead, as we observe in Figure \ref{fig:turboseti_rmse}, TurboSETI very accurately localizes these signals, to a median RMSE of about 1 -- 2 pixels for each SNR.}

For the ML predictions, there are still large outliers, even reaching levels of randomness (300 -- 500 pixels off) in the two signal case. One potential explanation and limitation in our model is that our labels are ordered. For two signals, at the highest SNR, both signals are at 25 dB. The model could have a harder time differentiating one signal for another (where the only distinguishing factor is zero vs. non-zero drift rate) and produce bad predictions. Our test set has 4,000 images for each SNR level, so even having a small fraction of these show up as outliers at the highest SNRs would be compounded if we tried using this model on real data. Recall that a $32\times1024$ px image at $1.4$ s and $1.4$ Hz resolution is about 45 s by 1430 Hz in total. For a typical 5 minute C-band observation (4 -- 8 GHz), this is equivalent to about 18.65 million data frames. Before it makes sense to use our CNN pipeline in searches on real data, we need to find ways of very precisely localizing these signals so that we are not swamped with false positives and inaccurate positions. 

Nevertheless, inspecting ML predictions in the two signal case revealed that the models appear to learn that the first two labels (i.e. corresponding to the starting and ending index of the zero drift RFI signal) should be the same index, since it predicted essentially the same value between [0, 1) up to a few significant figures (at least to differences of $1/1024$).

\newtext{Despite being much more accurate over the one signal dataset, TurboSETI takes much longer to produce localization predictions than our CNN models for the same amount of data, on the order of 20 minutes vs. 34 seconds, as discussed in Section \ref{subsec:turbo}. Of course, this is not necessarily surprising, since our ML predictions take advantage of GPU-accelerated calculations, especially when we batch together multiple data frame inputs at once. }

\newtext{For every frame in the two signal dataset, we also observe that TurboSETI struggles to find multiple signals that are close together or intersecting, and instead only detects the brighter one. For the same dataset, our final CNN model can generally localize such signals to an RMSE of about 10 -- 20 pixels for signals at 10 dB and above. Besides being more computationally efficient, CNN-based pipelines may therefore be better at finding elusive signals that standard search techniques tend to miss.}

In addition, it is encouraging that when we take into account the multiple possible normalizations, our model performance improves, especially in the two signal case with a model RFI signal. We believe that this could generalize well to frames with over two signals, as long as we can find a good way of matching labels and predictions, perhaps without necessarily enforcing an ordering.

\section{Summary}
\label{sec:summary}

Accurately identifying the presence and positions of signals in radio data is important for finding candidate technosignatures and ensuring that we do not miss interesting signals in the presence of bright RFI. Computer vision techniques allow us to ingest complex image frame data and distill them into relevant information, such as signal locations. 

We found that our final model outperformed our baseline model for all SNRs in both datasets, and training each model on datasets limited to 10 -- 25 dB results in significant increases in model performance. Our best results had a median RMSE of about 6 pixels in the one signal case and about 9 pixels in the two signal case. Since we used simple signals embedded in ideal chi-squared noise, we expected our localization models to perform even better, especially in the one signal case. Nevertheless, while these errors are higher than expected and come with a host of outliers that perform much more poorly, these results are promising for future work in localizing narrow-band signals in images. 

\newtext{We also did an analysis using TurboSETI to detect signals in our synthetic datasets, and compared the results with our ML predictions. We found that while TurboSETI produces more accurate localizations for the one signal dataset, our ML pipelines are much faster and able to produce meaningful results even for more complex tasks, such as localizing both signals in our two signal dataset. }

Overall, object detection and localization of long, thin objects is difficult since they do not match the typically rectangular shape of many other objects, and so it is harder to maximize intersection over union measures with ground truth. Although on the one hand detecting lines seems simple intuitively, the relative lack of information compared to broader, extended objects makes it more difficult, especially within a noisy background. Nevertheless, with a few key assumptions that are specific to the radio data we collect for SETI, we can make more progress in precisely localizing radio frequency signals. 

A future direction for this work is to investigate the effectiveness of treating multiple normalizations as independent inputs to the same model, instead of combining them as a single two channel input. Each normalization could have a few convolutional layers to itself, and would be added to each other to learn features with contributions from both normalizations. Indeed, this approach could scale better and benefit from additional data preprocessing techniques beyond the two normalizations discussed in this paper.

We can also easily use this CNN architecture to classify signals, or to both classify and localize simultaneously depending on how we choose our labels and loss functions. We are interested to see how well this method extends to more than two signals in a single image frame, and eventually, we would like to develop a pipeline for signal $\mathit{detection}$ of an arbitrary number of signals in a given image frame.

\section{Acknowledgements}

Breakthrough Listen is managed by the Breakthrough Initiatives, sponsored by the
Breakthrough Prize Foundation. The Green Bank Observatory is a facility of the National Science Foundation, operated under cooperative agreement by Associated Universities, Inc. We thank the staff at the Green Bank Observatory for their operational support.

\software{Keras \citep{chollet2015keras},
Setigen\footnote{\url{https://github.com/bbrzycki/setigen}},
TurboSETI \citep{enriquez2019turboseti},
Blimpy \citep{2019JOSS....4.1554P},
Matplotlib \citep{hunter2007matplotlib}}

\bibliographystyle{aasjournal}
\bibliography{references}

\end{document}